\makeatletter
\let\blx@rerun@biber\relax
\makeatother

\documentclass[runningheads]{llncs}
\usepackage{graphicx}
\usepackage[utf8]{inputenc}
\usepackage[english]{babel}

\usepackage[frozencache=true,cachedir=.]{minted}

\usepackage[sortcites=true,
    citestyle=numeric-comp,
    natbib=true,
    backend=biber,
    url=false,
    urldate=comp,
    maxbibnames=99]{biblatex}
\usepackage{hyperref}

\usepackage{algorithm} 
\usepackage{algpseudocode}
\algdef{SE}[DOWHILE]{Do}{doWhile}{\algorithmicdo}[1]{\algorithmicwhile\ #1}

\usepackage[operators,
            advantage,
            adversary,
            probability]{cryptocode}

\usepackage{wrapfig}

\begin{document}

\title{Towards Formal Verification of Password Generation Algorithms used in Password Managers\thanks{Supported by the PassCert project, a CMU Portugal Exploratory Project funded by Funda\c{c}\~ao para a Ci\^encia e Tecnologia (FCT), with reference CMU/TIC/0006/2019. Project URL: \protect\url{https://passcert-project.github.io}}}
%
%
\author{Miguel Grilo\inst{1} \and
João F. Ferreira\inst{2} \and
José Bacelar Almeida\inst{3}}
\authorrunning{M. Grilo et al.}
%
\institute{INESC TEC and IST, University of Lisbon, Portugal \and INESC-ID and IST, University of Lisbon, Portugal \and
HASLab, INESC TEC and University of Minho, Portugal
}
%
\maketitle

\begin{abstract}
Password managers are important tools that enable us to
use stronger passwords, freeing us from the cognitive burden of remembering them. Despite this, there are still many users who do not fully
trust password managers. In this paper, we focus on a feature that most
password managers offer that might impact the user’s trust, which is the
process of generating a random password. We survey which algorithms
are most commonly used and we propose a solution for a formally verified reference implementation of a password generation algorithm. We
use EasyCrypt as our framework to both specify the reference implementation and to prove its functional correctness and security.

\keywords{Password Manager  \and Random Password Generator \and Formal Verification \and Security}
\end{abstract}

\section{Introduction} \label{Introduction}
To address many of the existing problems and vulnerabilities regarding password authentication \cite{florencio2007large,zuo2019does,kamp2012linkedin,ferreira2017certified,johnson2020skeptic,pereira2020evaluating},
security experts recommend using password managers (PMs) for both storing and generating strong random passwords \cite{hunt2017passwords}. However, despite these recommendations, users tend to reject PMs, partly because they do not fully trust these applications \cite{alkaldi2016people,pearman2019people,aurigemma2017so}. It has been observed that generation of random passwords is one important feature that increases use of PMs \cite{alkaldi2016people} and helps prevent the use of weaker passwords and password reuse \cite{pearman2019people}. These studies suggest that a strong password generator that users can fully trust is a must-have feature for PMs.

In this paper, we propose a formally verified reference implementation for a Random Password Generator (RPG). We also present the formalization of functional correctness and security properties of our implementation, using the EasyCrypt proof environment \cite{barthe2013easycrypt} and the game-based approach for cryptographic security proofs \cite{shoup2004sequences, bellare2004code}.

\section{Current Password Generation Algorithms} \label{CurrentPasswordGenerationAlgorithms}
We studied 15 PMs to understand what are the most commonly used algorithms, but in this paper we focus on three of them: Google Chrome's PM (v89.0.4364.1)\footnote{https://source.chromium.org/chromium/chromium/src/+/master:components}, Bitwarden (v1.47.1)\footnote{https://github.com/bitwarden}, and KeePass (v2.46)\footnote{https://github.com/dlech/KeePass2.x}.
These were chosen because they are widely used and open-source, which allows us to access their source code and study them in detail.

\subsection{Password Composition Policies} \label{PasswordCompositionPolicies}

In general, PMs allow users to define password composition policies that the generated passwords must satisfy. These policies define the structure of the password, including the password's length and the different character classes that may be used. These policies are used to restrict the space of user-created passwords, thus precluding some easily guessed passwords.
Table \ref{tab:1} shows the policies that can be specified in the studied PMs.

\begin{table}
\scriptsize
\centering
\begin{tabular}{c|c|c|c|}
\cline{2-4}
                                                                                                                                          & \textbf{Chrome}                                                                                                           & \textbf{Bitwarden}                                                                                           & \textbf{KeePass}                                                                                                                                    \\ \hline
\multicolumn{1}{|c|}{\textbf{\begin{tabular}[c]{@{}c@{}}Password\\ Length\end{tabular}}}                                                  & 1-200                                                                                                                     & 5-128                                                                                                        & 1-30000                                                                                                                                             \\ \hline
\multicolumn{1}{|c|}{\textbf{Available sets}}                                                                                             & \begin{tabular}[c]{@{}c@{}}Lowercase Letters\\ Uppercase Letters\\ Alphabetic\\ Numbers\\ Special Characters\end{tabular} & \begin{tabular}[c]{@{}c@{}}Lowercase Letters\\ Uppercase Letters\\ Numbers\\ Special Characters\end{tabular} & \begin{tabular}[c]{@{}c@{}}Lowercase Letters\\ Uppercase Letters\\ Numbers\\ Special Characters\\ Brackets\\ Space\\ Minus\\ Underline\end{tabular} \\ \hline
\multicolumn{1}{|c|}{\textbf{\begin{tabular}[c]{@{}c@{}}Define minimum\\ and maximum \\ occurrence of\\ characters per set\end{tabular}}} & Yes                                                                                                                       & \begin{tabular}[c]{@{}c@{}}Yes. Can only\\ define minimum\end{tabular}                                       & No                                                                                                                                                  \\ \hline
\multicolumn{1}{|c|}{\textbf{\begin{tabular}[c]{@{}c@{}}Exclude similar \\ characters\end{tabular}}}                                      & Yes. \{l o I O 0 1\}                                                                                                      & Yes \{l I O 0 1\}                                                                                            & Yes \{l I O 0 1 $|$\}                                                                                                                                 \\ \hline
\multicolumn{1}{|c|}{\textbf{\begin{tabular}[c]{@{}c@{}}Define by hand\\ a character set\end{tabular}}}                                   & No                                                                                                                       & No                                                                                                           & Yes                                                                                                                                                 \\ \hline
\multicolumn{1}{|c|}{\textbf{\begin{tabular}[c]{@{}c@{}}Define by hand\\ a character  set\\ to be excluded\end{tabular}}}                 & No                                                                                                                        & No                                                                                                           & Yes                                                                                                                                                 \\ \hline
\multicolumn{1}{|c|}{\textbf{\begin{tabular}[c]{@{}c@{}}Remove \\ duplicates\end{tabular}}}                                               & No                                                                                                                        & No                                                                                                           & Yes                                                                                                                                                 \\ \hline
\end{tabular}
\medskip
\caption{\label{tab:1}Available policy options a user can define. The Alphabetic set in Chrome is the union of Lowercase Letters and Uppercase Letters. The set of Special Characters in Chrome and Bitwarden is \{- \_ . : !\} while in KeePass it is \{! " \# \$ \% \& ' * + , . / : ; = ? @ \textbackslash \^{} $|$\}. The Brackets set in KeePass is \{( ) \{ \} [ ] $\langle\rangle$\}. The Space, Minus, and Underline are the single element sets \{\textvisiblespace\}, \{-\}, and \{\_\}, respectively.}
\end{table}

\subsection{Random Password Generation} \label{RandomPasswordGeneration}
The main idea of the surveyed algorithms is to generate random characters from the different character sets until the password length is fulfilled, taking also into consideration the minimum and maximum occurrences of characters per set.
Chrome's algorithm starts by randomly generating characters from the sets which have the minimum number of occurrences defined. Then, it generates characters from the union of all sets which have not already reached their maximum number of occurrences. Lastly, it generates a permutation on the characters of the string, resulting in a random generated password.
Bitwarden's algorithm is similar, but it makes the permutation before generating the characters (i.e., it creates a string like `{\sl llunl}' to express that the first two characters are lowercase letters, followed by an uppercase letter, then a number, and finally a lowercase letter. Only then it generates the characters from the respective sets).
KeePass does not support defining the minimum and maximum occurrences of characters per set, so the algorithm just randomly generates characters from the union of the sets defined in the policy.

\subsection{String Permutation} \label{StringPermutation}
Given the need to generate a random permutation of the characters of a string, Bitwarden and Chrome both implement an algorithm to do so. The basic idea for both PMs is the same, which is to randomly choose one character from the original string for each position of the new string.

\subsection{Random Number Generator} \label{RandomNumberGenerator}
The RPG needs to have an implementation of a Random Number Generator (RNG) that generates random numbers within a range of values.
Chrome and KeePass use similar RNGs that generate numbers from 0 to an input \textit{range}. The main idea of these two PMs is to generate random bytes, then casting them to an integer, and then return that value modulo range, so the value it generates is between 0 and range. 
Bitwarden's RNG allows generating numbers from an arbitrary minimum value up to an arbitrary maximum value. Since in our algorithm we do not need this feature, the simpler implementations of Chrome and KeePass are sufficient for us to focus on.

All these RNGs call a random bytes generator. Regarding this, the three PMs considered use different approaches. Chrome uses system calls depending on the operating system it is running, Bitwarden uses the NodeJS \textit{randomBytes()} method, while KeePass defines its own random bytes generator based on ChaCha20.
In this work, we assume that random bytes generator exists and are secure.


\section{Reference Implementation} \label{ReferenceImplementation}
Based on our survey (Section \ref{CurrentPasswordGenerationAlgorithms}), we propose a reference implementation for an RPG which offers the following policy adjustments: (1) the user can define the password length (1-200); (2) the user can choose which sets to use (from Lowercase Letters, Uppercase Letters, Numbers, and Special Characters); (3) the user can define the minimum and maximum occurrences of characters per set. The restriction on the maximum length is just to avoid passwords that are unnecessarily large, since passwords with at least 16 characters are already hard to guess and secure \cite{shay2016designing}.
In Algorithm \ref{alg:4} we show the pseudo-code of the proposed reference implementation.

Our algorithm receives as input the password composition policy. Then it randomly generates characters from the sets that have a \textit{min} value greater than 0, and appends them to the \textit{password} (initially an empty string). Then, until the size of \textit{password} is equal to the length defined in the policy, it randomly generates characters from the union of all sets which have not more than their \textit{max} value of characters in \textit{password}. Finally, it generates a random permutation of the string, and returns it.

\begin{algorithm}
	\caption{RPG Reference Implementation} 
	\begin{algorithmic}[1]
	    \Procedure{Generate}{$policiy$}
    	    \State $pwLength \leftarrow policy.pwLength$
    	    \State $charSets \leftarrow policy.charSets$
    	    \State $password \leftarrow \varepsilon$
    		\ForAll {$set\in{charSets}$}
    			\For {$i=1,2,\ldots,set.min$}
    				\State $char \leftarrow \textsc{GenerateCharacter}(set)$
    				\State $password \leftarrow password || char$
    			\EndFor
    		\EndFor
    		\While {$len(password) < pwLength$}
    		    \State $availableSets \leftarrow \bigcup_{set\in{charSets}} set$ such that $set.max > 0$
    		    \State $char \leftarrow \textsc{GenerateCharacter}(availableSets)$
    		    \State $password \leftarrow password||char$
    		\EndWhile
    		\State $password \leftarrow \textsc{Permutation}(password)$
    		\State \Return $password$
    	\EndProcedure
    	\State
    	\Procedure{GenerateCharacter}{$set$}
    	\State $choice \leftarrow {RNG}(set.size)$
    	\State $set.max \leftarrow set.max - 1$
    	\State \Return $choice$
    	\EndProcedure
    	\State
        \Procedure{Permutation}{$string$}
            \For {$i=len(string)-1,\ldots,0$}
    		    \State $j \leftarrow {RNG}(i)$
    			\State $aux = string[i]$
    			\State $string[i] = string[j]$
    			\State $string[j] = aux$
    		\EndFor
    		\State \Return $string$
    	\EndProcedure
    	\State
        \Procedure{RNG}{$range$}
	        \State $maxValue \leftarrow (uint64.maxValue / range) * range - 1$
    		\Do
                \State $value \leftarrow$ (uint64) GenerateRandomBytes
            \doWhile{$value > maxValue$}
            \State \Return $value\ \textrm{mod}\ range$
    	\EndProcedure
	\end{algorithmic}
\label{alg:4}
\end{algorithm}

\section{Formal Proofs} \label{Formal Proofs}
In this section we present our two main properties to be proved about our RPG: functional correctness and security.

\subsection{Functional Correctness} \label{Functional Correctness}
We say that an RPG is functionally correct if generated passwords satisfy the input policy (the probability that the generated password will satisfy the policy is 1).
This property guarantees that users will always get an output according to their expectations. In the following code we show a succinct version of the formalization of functional correctness in EasyCrypt style.
\begin{minted}{haskell}
module Correctness(RPG : RPG_T) = {

    proc main(policy:policy) : bool = {
        var password : password;
        var satLength, satBounds : bool;
        password <@ RPG.generate_password(policy);
        satLength <@ satisfies_length(password, policy);
        satBounds <@ satisfies_bounds(password, policy);
        return satLength /\ satBounds;
    }
}.
\end{minted}
The procedures \mintinline{haskell}{satisfies_length} and \mintinline{haskell}{satisfies_bounds} check, respectively, if the password's length is the same as the one defined in the policy and if the \textit{max} and \textit{min} values defined per set in the policy are satisfied.

Using this definition, we can define the lemma \mintinline{haskell}{correctness_phl} which can be proved using probabilistic Hoare logic (pHL):

\begin{minted}{haskell}
lemma correctness_phl (p:policy) :
  Pr[Correctness(RPGRef).main : policy = p ==> res] = 1%r.
\end{minted}

\subsection{Security} \label{Security}
\setlength\intextsep{0pt}
\begin{wrapfigure}{r}{0.4\textwidth}
\begin{pcvstack}[boxed, center]
\footnotesize
    \begin{pchstack}[center, space=1em]
        \procedure[mode=text]{Game Real$_\adv$()}{
        $b \leftarrow$ \adv\textsuperscript{RealRPG(.)}() \\
        return $b$
        }
    \end{pchstack}
    \begin{pchstack}[center, space=1em]
    \procedure[mode=text]{Game Ideal$_\adv$()}{
        $b \leftarrow$ \adv\textsuperscript{IdealRPG(.)}() \\
        return $b$
        }
    \end{pchstack}
    
    \begin{pchstack}[center, space=1em]
        \procedure[mode=text]{proc RealRPG(policy)}{
        return RPGRef.generate(policy)
        }
    \end{pchstack}
    \begin{pchstack}[center, space=1em]
         \procedure[mode=text]{proc IdealRPG(policy)}{
        $password \sample p \subset P$ \\
        return $password$ 
        }
    \end{pchstack}
\end{pcvstack}
  \caption{Security games. RPGRef is our reference implementation and $p$ is the subset of the set of all possible passwords $P$ that satisfy the given policy.}
  \label{fig:secgames}
\end{wrapfigure}

Regarding security, we want to verify that, given the policy, the generated password has the same probability of being generated as any other possible password. In other words, considering the entire set of possible passwords defined by the policies, we want to make sure that there is a uniform distribution over that set. To prove such property we can use the game-based approach for cryptographic security proofs \cite{shoup2004sequences, bellare2004code}. This methodology is supported by EasyCrypt.

The notion of security for an RPG is expressed by two games Real$_\adv$() and Ideal$_\adv$(), which are parameterized by an attacker \adv. This attacker has access to an oracle in each game (one for our reference implementation and one for the ideal RPG). The attacker is trying to guess with which oracle he is interacting with, by outputting a boolean \textit{b}. The games are shown in Figure \ref{fig:secgames}.

In order to consider our implementation secure, the attacker can not be able do distinguish between it and the ideal RPG. This means that the following advantage must be neglibible

\begin{center}
    \advantage{RPG}{}[(\adv)] = Pr[Real$_\adv$() $\Rightarrow$ res] - Pr[Ideal$_\adv$() $\Rightarrow$ res]
\end{center}
where Pr[\textit{game}() $\Rightarrow$ res] is the probability that \textit{game}() outputs true.

\section{Conclusion}
In this paper we presented an analysis of the RPG algorithms currently used in popular PMs, and we proposed a reference implementation.
Then, we showed the formalization of two properties that we can verify using EasyCrypt: its functional correctness and security.

As future work, after proving the properties, we will implement the reference using Jasmin \cite{almeida2017jasmin}, a framework for developing high-speed and high-assurance cryptographic software.
We might also extend the policy composition options with, for example, the option for the user to define a character set ``by hand''.
While generally speaking strict password composition policies are a good security mechanism, these can still generate some easily guessed passwords (e.g., a policy that enforces the use of  all character classes may generate the easily guessed password ``P@ssw0rd'') \cite{shay2016designing}. So, it might also be interesting to define and formalize a property regarding password strength, which would guarantee that our RPG would only generate strong passwords
(according to some metric).


%
%
\printbibliography

\newpage
\appendix
\section{Appendix: Algorithms Discussed in Section \ref{CurrentPasswordGenerationAlgorithms}}

\subsection{RPG Algorithm}
Following the discussion in Section~\ref{CurrentPasswordGenerationAlgorithms},
we present the following pseudo-code (Algorithm \ref{alg:1}) which is a generalization of the algorithms of the three password managers, being closely similar to Chrome's algorithm.

\medskip

\begin{algorithm}
	\caption{General Password Generation Algorithm} 
	\begin{algorithmic}[1]
		\Procedure{GeneralGenerate}{$policies$}
    	    \State $pwLength \leftarrow policies.pwLength$
    	    \State $charSets \leftarrow policies.charSets$
    	    \State $password \leftarrow \varepsilon$
    		\ForAll {$set\in{charSets}$}
    			\For {$i=1,2,\ldots,set.min$} \Comment{set.min is the minimum number of characters from that set that must appear on the generated password}
    				\State $char \leftarrow$ Randomly generate a char from the set
    				\State $password \leftarrow password || char$ \Comment{`$||$' is the append operator}
    			\EndFor
    		\EndFor
    		\While {$len(password) < pwLength$}
    		    \State $char \leftarrow$ Randomly generate a char from the union of all sets, while not exceeding the $max$ char occurrence of each set
    		    \State $password \leftarrow password || char$
    		\EndWhile
    		\State Generate a random permutation of $password$
    		\State Output $password$
    	\EndProcedure
	\end{algorithmic}
\label{alg:1}
\end{algorithm}

\subsection{String Permutation} 
The method described in Section~\ref{StringPermutation}
is described in pseudo-code in Algorithm \ref{alg:2}.

\medskip

\begin{algorithm}
	\caption{String Permutation} 
	\begin{algorithmic}[1]
	    \Procedure{GeneralPermutation}{$string$}
            \For {$i=len(string)-1,\ldots,0$}
    		    \State $j \leftarrow$ Random number between 0 and i
    			\State $aux = string[i]$
    			\State $string[i] = string[j]$
    			\State $string[j] = aux$
    		\EndFor
    	\EndProcedure
	\end{algorithmic}
\label{alg:2}
\end{algorithm}

\subsection{Random Number Generator} 
Chrome and KeePass RNG algorithms 
described in Section~\ref{RandomNumberGenerator}
are described in pseudo-code in Algorithm \ref{alg:3}.

\begin{algorithm}
	\caption{RNGs with maximum range} 
	\begin{algorithmic}[1]
	    \Procedure{ChromeRNG}{$range$}
	        \State $maxValue \leftarrow (uint64.maxValue / range) * range - 1$ \Comment{uint64.maxValue is the maximum value an unsigned integer with 64 bits can take; `/' is the Euclidean division}
    		\Do
                \State $value \leftarrow$ (uint64) GenerateRandomBytes
            \doWhile{$value > maxValue$}
            \State \Return $value\ \textrm{mod}\ range$
    	\EndProcedure
    	\State
	    \Procedure{KeePassRNG}{$range$}
    		\Do
                \State $genValue \leftarrow$ (uint64) GenerateRandomBytes
                \State $value \leftarrow genValue\ \textrm{mod}\ range$
            \doWhile{$(genValue - value) > (uint64.maxValue - (range - 1))$}
            \State \Return $value$
    	\EndProcedure
	\end{algorithmic}
\label{alg:3}
\end{algorithm}

\end{document}